# Molecular chirality and the orbital angular momentum of light

D. L. Andrews[†*], L. C. Dávila Romero[†] and M. Babiker[§]

[†]*School of Chemical Sciences and Pharmacy, University of East Anglia, Norwich, NR4 7TJ, U.K.*

[§]*Department of Physics, University of York, Heslington, York YO10 5DD, U.K.*

Abstract

Optical beams with a new and distinctive type of helicity have become the subject of much recent interest. While circularly polarised light comprises photons with *spin* angular momentum, these optically engineered 'twisted beams' (*optical vortices*) are endowed with *orbital* angular momentum. Here, the wave-front surface of the electromagnetic fields assumes helical form. To date, optical vortices have generally been studied only in their interactions with achiral matter. This study assesses what new features, if any, can be expected when such beams are used to interrogate a chiral system.

1. **Background**

Molecular chirality signifies a structural handedness associated with variance under spatial inversion or a combination of inversion and rotation, equivalent to the usually stated criterion of a lack of any improper axes of rotation. Traditionally, chiral optics engages circularly polarized light – even in the case of optical rotation, interpretation of the phenomenon commonly requires the plane polarized state to be understood as a superposition of circular polarizations with opposite handedness. For circularly polarized light, the left- and right- forms designate the sign of intrinsic spin angular momentum, $\pm h$, and also the helicity of the locus described by the associated electromagnetic field vectors. For this reason its interactions with matter are enantiomerically specific.

The considerable interest in optical angular momentum has been enhanced through realisation of the possibility to engineer optical vortices [1-5]. Here, helicity is present in the wave-front surface of the electromagnetic fields and the associated angular momentum is termed 'orbital'. The radiation itself is commonly referred to as a 'twisted' or 'helical' beam. Mostly, optical vortices have been studied only in their interactions with achiral matter – the only apparent exception is some recent work on liquid crystals [6-8]. It is timely and of interest to assess what new features, if any, can be expected if such beams are used to interrogate any system whose optical response is associated with enantiomerically specific molecules. Section 2 introduces salient quantum electrodynamical principles for discussion, in section 3, of issues relating to the interplay of quantised spin and orbital angular momentum – the latter with particular reference to twisted beams. The key results are summarised in section 4.

## 2. Fundamental principles of molecular chirality

We first have to construct in generalised form the criteria for manifestations of chirality in optical interactions. For simplicity, materials with a unique enantiomeric specificity are assumed – signifying a chirality that is intrinsic and common to all molecular components (or chromophores) involved in the optical response. Results for systems of this kind will also apply to single-molecule studies. Longer-range translation/rotation order can also produce chirality, as for example in twisted nematic crystals, but such mesoscopic chirality cannot directly engender enantiomerically specific interactions. The only exception is where optical waves probe two or more electronically distinct, dissymmetrically oriented but intrinsically achiral molecules or chromophores (see below).

Chiroptical interactions can be distinguished by their electromagnetic origins: for molecular systems in their usual singlet electronic ground state, they involve the spatial variation of the electric and magnetic fields associated with the input of optical radiation. This variation over space can be understood to engage chirality either through its coupling with dissymmetrically placed, neighbouring chromophore groups (Kirkwood's two-group model [9], of limited application) or more generally through the coupling of its associated electric and magnetic fields with individual groups. As chirality signifies a local breaking of parity it permits an interference of electric and magnetic interactions. (Even in the two-group case, the paired electric interactions of the system corespond to electric and magnetic interactions of the single entity which the two groups comprise. Thus, for convenience, the term 'chiral centre' is used in the following to denote either chromophore or molecule.

**2.1 Quantum formulation**

In the framework of quantum electrodynamics, the Hamiltonian comprises the unperturbed operators for the chiral centre and the radiation, and also the interaction Hamiltonian, whose role as perturbation operator leads to optical transitions. Each component of the Hamiltonian is necessarily of even parity with respect to space inversion and also even with respect to time reversal. The interaction term is expressible in either minimal coupling form (in terms of the vector potential of the radiation field) or the more familiar multipolar formulation, directly expressed in terms of electric and magnetic fields. These two (and other) options lead to identical results for real processes, that is those subject to overall energy conservation [10, 11]; for convenience we use the multipolar form.

The multipolar interaction Hamiltonian entails a linear coupling of the molecular polarisation field (accommodating all electric multipoles E$n$) with the transverse electric field $\mathbf{e}(\mathbf{r})$ of the radiation, and also a linear coupling of the molecular magnetisation field (all magnetic multipoles M$n$) with the magnetic field $\mathbf{b}(\mathbf{r})$ of the radiation. For present purposes we restrict consideration to the E1 and M1 interactions whose interplay is mostly associated with manifestations of chirality, leaving the electric quadrupole and other higher multipoles for detailed appraisal in a following piece of work. The interaction Hamiltonian for a system of molecules labelled *x* is thus expressible as;

$$H_{int} = -\sum_{x} \boldsymbol{\mu}(\mathbf{r}_x) \cdot \mathbf{e}(\mathbf{r}_x) - \sum_{x} \mathbf{m}(\mathbf{r}_x) \cdot \mathbf{b}(\mathbf{r}_x). \tag{1}$$

The electric dipoles $\mathbf{m}(\mathbf{r})$ have an odd signature for space parity and even for time; the magnetic dipoles $\mathbf{m}(\mathbf{r})$ have the opposite, even for space and odd for time. Equally, the electric field $\mathbf{e}(\mathbf{r})$ of the radiation field is space-odd and time-even, while the magnetic field $\mathbf{b}(\mathbf{r})$ is the opposite. Note, however, that the latter assertions are true *only* for the electric and magnetic fields of the radiation as a whole, *not* necessarily when only one or a finite number of modes is considered (independent of the mode expansion used). These and other fundamental symmetry properties are summarised in Table 1.

The quantum amplitude $M_{fi}$ for a specific optical interaction in a single chiral centre is constructed from time-dependent perturbation theory and entails a linear combination of scalars, each of which is the inner product of two rank $r$ tensors: $\mathbf{S}^{(r)}$, comprising an outer product of radiation components (specifically, a product of components of the electric field and magnetic field) and $\mathbf{T}^{(r)}$, an outer product of molecular multipole components [12];

$$M_{fi}^{x} = \exp(i\Delta\mathbf{k}\cdot\mathbf{R}_{x}) \sum_{e,m=0}^{n} \mathbf{S}^{(r)}_{e;\,m;\,n-e-m} \otimes \mathbf{T}^{(r)}_{e;\,m;\,n-e-m} . \qquad (2)$$

Here, ?$\mathbf{k}$ is the mismatch between the wave-vector sum of all input and the sum of all output photons (if any) involved in the process at a chiral centre $x$ located at $\mathbf{R}_x$. The molecular tensor $\mathbf{T}^{(r)}$ can be written as a product of one or more molecular transition integrals, determined by the number of photons involved. In every case the origin of such forms is a single transition integral, expressible as a Dirac bracket with the appropriate number of electromagnetic interactions coupling the initial state to the final state. In order for the transition integral not to vanish identically, the triple product of the group theoretic representations for the initial and final state wavefunctions with that of the interaction must contain the totally symmetric representation.

Before considering chiral systems, consider the case of a centrosymmetric molecular system. Here, the only non-zero contributions to the transition integral can be those whose parity equates to the product signature of the initial and final state wavefunctions. For example if electric dipole coupling is to give a non-vanishing contribution then, for that transition, electric quadrupole and magnetic dipole contributions vanish by virtue of their opposite spatial parity. Thus, for centrosymmetric species, all multipolar contributions to any given quantum amplitude will have the same spatial parity. Clearly for chiral systems, where parity is not a good quantum number, no such rule applies and the amplitude may entail contributions of both positive and negative parity. In calculating the observable associated with a particular optical interaction it is usual to apply the Fermi Rule,

$$\Gamma \propto \left| M_{fi} \right|^2 = \sum_{\boldsymbol{x},\boldsymbol{x}'}^{N} M_{fi}^{\boldsymbol{x}} \bar{M}_{fi}^{\boldsymbol{x}'} . \tag{3}$$

For chiral species this includes, in addition to diagonal terms that are all of even parity in both radiation and molecular parameters, terms associated with the quantum interference of couplings with opposite parity, of odd parity for both the radiation and molecule.

**2.2 System response**

Not only molecular symmetry determines the nature and extent of any chiroptical response; macroscopic symmetry is important too. In particular, when the system of interest is any fluid or other microscopically disordered medium, the isotropic symmetry of the bulk comes into play. In general, measurements result from the optical interactions of more than one chiral centre, and the observable signal results from a quantum amplitude comprising an ensemble sum of contributions. As is apparent from the $\exp(i\Delta \mathbf{k} \cdot \mathbf{R})$ factor in (2), each of these contributions generally has a different phase associated with spatial variance in the registration of light at each centre. For most optical processes, which are *non-parametric* (*i.e.* $\Delta \mathbf{k} \neq 0$), the interference of quantum amplitudes from different centres gives a null contribution when those centres are isotropically distributed – except in certain nanomaterials [13,14]. For such incoherent processes, no role is played by any orientational order characterising *long*-range chirality. Important exceptions arise for *parametric* optical processes ($\Delta \mathbf{k} = 0$). In the following, attention is focused on incoherent processes for two reasons; (i) it is processes of this category that are involved in the most common manifestations of chirality (circular dichroism, differential Rayleigh and Raman scattering, circularly polarised luminescence etc.; (ii) where optical vortices are concerned, it is known that parametric processes generally entail conservation of orbital angular momentum by the radiation field, such that chirality is not engaged [15].

**2.3 Incoherent processes; quantum interference terms**

We now focus specifically on chiroptical observables associated with incoherent processes in which the ensemble represents the time-averaged response of a single chiral centre. For any such molecule the transition rate is a sum of amplitude contribution products, featuring amongst which are quantum interferences which are odd in parity for both the matter and radiation – which can only arise in the case of chiral systems. Handedness is apparent in two respects. If the space inversion operator (signifying a change to the opposite enantiomeric form) is applied to all properties of the molecule – and in particular its multipole moments – but not the radiation (signifying retention of its circularity), these interference terms change sign. The same is true if the radiation changes handedness but the molecule retains its enantiomeric form: all contributions to the signal are invariant to inversion of the whole system. This is why chiral interactions must involve handed radiation. Plane polarisations are invariant under space inversion; consequently, applying space inversion to a molecular system interacting with plane polarised light has the same effect as applying it to both the molecule and the radiation, and no chiral specificity can emerge. Optical rotation, as noted earlier, is an exception since it is a parametric process.

For the interference terms that support chiral selectivity, further conditions need to be satisfied to ensure that they are not identically zero. For any component of a non-parametric signal – and in particular the odd-parity quantum interference terms – rotational averaging effects a disentanglement of the radiation and molecular fields, $\mathbf{S}^{(n)}$ and $\mathbf{T}^{(n)}$ respectively. Rotational averaging results in a product of scalars (or, for tensors of rank four or more, a linear combination of such scalars), one scalar for the radiation and one for the molecule. Each scalar is derived by contracting the tensor, $\mathbf{S}^{(n)}$ or $\mathbf{T}^{(n)}$, with an isotropic tensor of the same rank. For example in the E1-M1 interference term for photon absorption by a chiral molecule the molecular tensor $\mathbf{T}^{(2)} = [\mathbf{m}][\mathbf{m}^*]$, on contraction with the isotropic tensor of rank 2 (the Kronecker delta), yields the scalar $T = (\mathbf{m} \cdot \mathbf{m}^*)$ – signifying that the electric and magnetic transition moments must not be orthogonal (in transitions between non-degenerate states, $\mathbf{m}^*$ will be antiparallel to $\mathbf{m}$). Equally $\mathbf{S}^{(2)} = [\mathbf{e}][\mathbf{b}^*]$ yields the scalar $S = (\mathbf{e} \cdot \mathbf{b}^*)$; this determines that chiral resolution vanishes when plane polarised light is employed, for then the field polarisations are real and the orthogonality of the electric and magnetic vectors gives $(\mathbf{e} \cdot \mathbf{b}^*) = 0$. But for any circular (or even for elliptical) polarisations, $\mathbf{S}$ is non-zero and, since it also changes sign on reversal of circularity;

$$S = (\mathbf{e} \cdot \mathbf{b}^*) \xrightarrow{I} ((-\mathbf{e}) \cdot \mathbf{b}^*) = -S \ , \qquad (4)$$

chiral specificity is manifest. Since $T$ takes opposite signs for different enantiomers,

$$T = (\mathbf{m} \, \mathbf{m}^*) \xrightarrow{\mathscr{I}} ((-\mathbf{m}) \cdot \mathbf{m}^*) = -T \ , \qquad (5)$$

the rate of absorption of left-handed circularly polarised radiation, for example, is different for left- and right-handed enantiomers – though only marginally, because of the relative weakness of the salient interference terms compared to the dominant (usually electric dipole) diagonal contributions to the signal. Equally, each enantiomer exhibits a slightly different rate of absorption for left and for right-handed circular input. This is the origin of circular dichroism. Similar remarks apply to optically more intricate processes involving more than one photon, where the quantum amplitude itself comprises products of multipolar couplings, each with a definite resultant spatial parity, and the key interference terms arise from products of these products with opposite inversion symmetry. Again, for a chirally specific signal to emerge, it is necessary that the resultant radiation and molecular scalars do not vanish identically.

## 3. The interactions of twisted beams

We now consider by what means, if any, the chirality of a twisted beam can engage with that of matter. The helicity of optical vortices, which is present in their wave-front structure, is also manifest through additional phase factors in the positive and negative frequency components. In particular, the analytic signal for the electric field of a Laguerre-Gaussian (LG) mode propagating in the $z$-direction is [16];

$$\mathbf{e}(\mathbf{r}) = \hat{\mathbf{e}} f_{lp}(r) \exp\left[i(kz - l\boldsymbol{j})\right] ,  \qquad (6)$$

where $\hat{\mathbf{e}}$ is the polarization vector (for either plane or circularly polarised light, in the latter case taking the form $\hat{\mathbf{e}}^{(L/R)} = \frac{1}{\sqrt{2}}\left(\hat{\mathbf{i}} \pm i\hat{\mathbf{j}}\right)$ for the complex vector) and $f_{lp}(r)$ is a radial distribution function; $l$ is the orbital angular momentum quantum number, for which a positive sign denotes left helicity and a negative sign, right [17]. Of the two space-dependent terms in the field phase, the first is the origin of the wave-vector mismatch factor in (2), and the second is an azimuthal phase factor characteristic of twisted beams. The first component of the phase in (6) changes sign on space inversion (and on complex conjugation) just as the circular polarisation vectors behave. The second phase factor changes by addition of $\boldsymbol{p}$ on space inversion. The magnetic field for a twisted beam has an exactly similar form [18].

No difference should be expected between the behaviour of right and left forms of any optical vortex if the photons it comprises are plane polarised and the material they interact with is achiral, in accordance with the dictates of parity. This is equivalent to the case of circularly polarised, planar wave-front (e.g. Gaussian) modes, interacting with achiral material. However, there are several differences between LG photons and those associated with planar wave-fronts. In the present context a significant difference is that plane waves have no restriction in the direction of propagation, while the symmetry of LG photons designates propagation in one direction, $\hat{\mathbf{z}}$. This indicates that mirror inversion along a plane containing the $\hat{\mathbf{z}}$ axis becomes the relevant symmetry element. Arguing symmetry on these grounds is sufficient to establish that there is no differentiation between LG beams of opposite handedness when achiral material is interrogated.

The interaction of an optical vortex with chiral matter affords up to eight possibilities for combinations of different handedness; each molecule, each photon and the vortex itself can be of either handedness. Consider first the simpler case of a vortex beam comprising plane polarised photons, superficially analogous to the case of circular polarisations in a plane wave. Mirror inversion of either part of the system (molecule or radiation) gives a system different from the original, and a change in signal signifying chiral specificity might be anticipated. Representing in brackets the circularity of the radiation and the matter respectively, both cases require identity of (*L*, *L*) and (*R*, *R*), and so too (*L*, *R*) and (*R*, *L*). However (*L*, *L*) and (*L*, *R*) differ. More succinctly, denoting either handedness *L*/*R* by a circularity $c$ and the reverse by $c^*$, we have $(c,c) = (c^*,c^*)$ which may or may not be equal to $(c,c^*)$.

For the twisted beam case it is therefore necessary to ascertain whether there is a mechanism, by means of which signals that are *permissibly* different in these terms may *in fact* differ. In the previous section it was seen that in a fluid, the observable is obtained as the ensemble sum of contributions from all relevant centres. As in the case of plane waves where the factor $\exp(i\Delta \mathbf{k} \cdot \mathbf{r}_x)$ plays an important role, a similar factor, $\exp(-i\Delta l \boldsymbol{j})$ enters the radiation tensor $\mathbf{S}^{(n)}$; here $\Delta l$ is the mismatch between the orbital angular momentum sum of all input and the sum of all output photons (if any) involved in the process at a chiral centre $\boldsymbol{x}$ located at $\mathbf{r}_x = (r\boldsymbol{j}, z)$. If the optical process is parametric, $\Delta l = 0$ and all dependence on the sign of the orbital angular momentum of the radiation is lost. The transition rate, proportional to $\left|\langle M_{fi}\rangle\right|^2$, has no dependence on the sign of the orbital angular momentum $l$ and therefore is not a chiral discriminator. For a *non*-parametric process where $\Delta l \neq 0$, the incoherent character means that the rate is proportional to $\left\langle \left|M_{fi}\right|^2 \right\rangle$, in which the azimuthal phase factor cancels; again the transition rate does not involve the orbital angular momentum or its sign.

Finally we address the case of chiral molecules interacting with twisted light comprising circular photons. In view of the observations made above, the problem reduces to exactly that which applies to other beams of circular polarisation – but now there is a mechanism for the associated signals to actually differ. Again the helicity of the wavefront is immaterial so far as chiral specificity is concerned. Consider for example a left-handed twisted beam comprising left-handed circular photons interacting with a system of left-handed molecular enantiomers – and the simplest optical process in which chirality can be manifest, namely circular dichroism. Then, the quantum interference terms responsible for chiral discrimination change sign if either the circularity of the photons or the isomeric form of the molecules is changed; the interference terms are invariant to a change of both. No component of the rate changes if, for example, the beam becomes a right-handed vortex but the sense of the photons and molecular handedness is the same as before. Note that this conclusion is contingent upon the azimuthal phase factors for the electric and magnetic fields being identical; however this is a necessary consequence of Maxwell's equations, as has been shown explicitly for quantised Laguerre-Gaussian modes [15].

## 4. Conclusion

Through analysis of the grounds for chiroptical behaviour involving electric and magnetic dipole interactions it has been established that the helicity of optical vortices cannot engage through any parametric or non-parametric optical process with the chirality of a molecular system, other than through circularity of its photons. Thus, the manifestations of orbital angular momentum differ markedly from those associated with photon spin angular momentum. Preliminary work on circular differential scattering suggests that our conclusions remain valid even when electric quadrupole interactions are taken into account. Assessing the possible chiral selectivity of the *mechanical* interactions of such twisted beams is a separate issue which merits separate attention.

We conclude by noting one exceptional case, where certain of the above assumptions are unwarranted, concerning second harmonic generation by reflection at a chiral surface. This system has long-range orientational order, and local chirality is modified as the space within which surface molecules reside is no longer isotropic. The process has parametric character, because surface reflection satisfies a reduced requirement for wave-vector matching – one that applies only to the surface-parallel components of the two input and single output photon wave-vectors. As has been shown elsewhere [19, 20], a chirally sensitive response then emerges through electric dipole coupling alone. The characteristics of such a process with a twisted beam input are as yet to receive attention – the system presents considerable experimental and theoretical challenges.

**Acknowledgments**

We gratefully acknowledge financial support from the Engineering and Physical Sciences Research Council.

| Operator | Time parity | Space parity |
|---|---|---|
| Hamiltonian | +1 | +1 |
| Electric field | +1 | −1 |
| Electric field of LG mode $(k, \mathbf{l}, l, p)$ | +1 | $(-1)^{l+1}$ |
| Electric dipole | +1 | −1 |
| Magnetic field | −1 | +1 |
| Magnetic field of LG mode $(k, \mathbf{l}, l, p)$ | −1 | $(-1)^{l}$ |
| Magnetic dipole | −1 | +1 |
| Wave-vector | −1 | −1 |

**Table 1:** The space and time parity of different physical quantities.